\documentclass[iop]{emulateapj-rtx4} 
\shortauthors{Sekanina}
\shorttitle{2I/Borisov in HST Image on October 12}
\slugcomment{Version \today }

\begin{document}
\title{Note on the HST Image of Comet 2I/Borisov Taken on 2019 October 12}
\author{Zdenek Sekanina}
\affil{Jet Propulsion Laboratory, California Institute of Technology,
  4800 Oak Grove Drive, Pasadena, CA 91109, U.S.A.}
\email{Zdenek.Sekanina@jpl.nasa.gov.}

\begin{abstract}
The image of comet 2I/Borisov taken by the Hubble Space Telescope on October~12
shows a fan-like feature in the inner coma, which extends over a few arcseconds
from the nuclear condensation, its axis pointing a little west of the north.  It
is proposed that this protrusion is a halo of centimeter-sized and larger dust
grains or pebbles released at low velocities from the nucleus at large
heliocentric distances (on the order of 10~AU or more) before perihelion and
subjected to very low radiation pressure accelerations.
Unlike potential microscopic dust from ongoing activity, this
population of large grains should show no silicate feature near 10~$\mu$m in the
infrared and its presence in 2I/Borisov can be tested observationally.  The
existence of material of a similar size range in Oort Cloud comets
with perihelia near or beyond the snow line has long been documented by the
characteristic orientation of their tails.
\end{abstract}


\section{Introduction}
An interesting characteristic of the broadly disseminated image of
2I/Borisov, taken by the WFC3/UVIS camera  on board the Hubble
Space Telescope~(HST)~on October 12 when the comet was 2.79~AU
from Earth (implying a projected distance of about 2000~km
per arcsecond) and 2.37~AU from the Sun, is a prominent
dichotomy between the inner and outer regions of the comet's
undoubtedly dust dominated atmosphere.  While the outer
coma, developing gradually into an incipient tail at
distances of several tens of thousands of kilometers from
the nucleus, is extending toward the northwest, the most
conspicuous feature in the inner coma, at less than 10\,000~km
from the nucleus, is a sector of material emanating from the
nuclear condensation in a direction that is much closer to the north
than the tail.  Cursory inspection suggests that this protrusion
has a fairly sharp boundary on its eastern (sunward) side, but
the axis of its maximum extension deviates from the north a little
to the west; the bright part does not extend farther in projection
onto the plane of the sky than several thousand kilometers from
the nucleus.

\begin{table}[hb]
\vspace{-4.2cm} 
\hspace{4.54cm}
\centerline{
\scalebox{1.07}{
\includegraphics{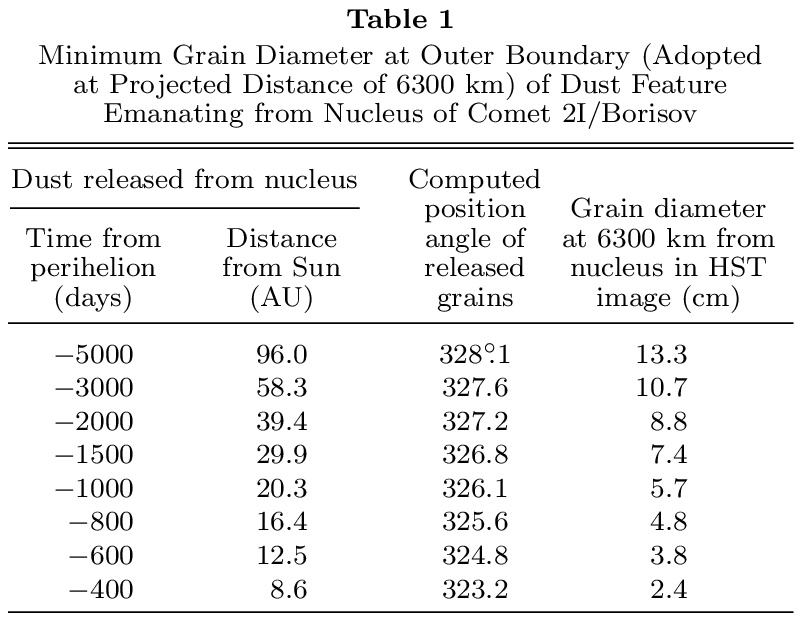}}}
\vspace{-20.85cm} 
\end{table}

\section{Discussion and Conclusions}

One could be tempted to interpret this bright feature as a manifestation
of ongoing activity, if the prevailing direction of the feature did not
make~an~\mbox{angle}~of~more than 120$^\circ$ with the projected direction to
the~Sun.~Accordingly, a more plausible explanation
is offered by the~apparent correspondence between the positions
of the fan's axis and the negative orbital velocity vector
whose predicted position angle is 330$^\circ$ (Sekanina 2019).
The negative orbital velocity vector is known to be the
direction of motion, relative to the nucleus, of dust grains
released from the comet at very early times before perihelion,
when the comet was at considerable heliocentric distances.  However,
in order for the grains to stay within a few thousand kilometers
of the nucleus over extraordinarily long periods of time,~it~is~essential~that
both the velocities of release~and~the~radiation~pressure
accelerations that the grains have been subjected to,~be {\it extremely\/}
low.  The momentum by loss~of~ice~and~random effects should have
broadened the initially filament-like feature over the long time; for example, a
halo of released grains expanding chaotically at a typical rate of, say,
0.1~m~s$^{-1}$ should occupy a volume of space 6300~km in extent
after two years, implying the separation from the nucleus
at 15~AU from the Sun.  To make certain that the magnitude of the integrated
radiation pressure effect on the grains over this period of time is
not greater, the grains should be subjected to
accelerations not exceeding $\sim$0.005~percent of the Sun's gravitational
acceleration.  Such grains should in fact be pebbles
at least \mbox{4--5 cm} across, if their scattering efficiency for
radiation pressure is unity and their bulk density equals
$\sim$0.5~g~cm$^{-3}$; the grain diameter varies as the
scattering efficiency and inversely as the bulk density.
It also varies inversely as the distance from the nucleus,
so that the largest pieces of this dust population are in
the nuclear condensation itself.  Similar grain dimensions
are obtained for a wide range of heliocentric distances
at the time of release, as presented in Table~1.~It is
possible~to~observationally distinguish between this population
of large grains~and ongoing activity if it should involve microscopic
dust by~testing for the 10-$\mu$m~silicate feature in the
infrared.  Evidence for icy-dust grains of similar
size released at very large heliocentric distances
prior to perihelion from the nuclei
of Oort Cloud comets with perihelia near or beyond the snow
line is provided by the diagnostic orientation of these
comets' tails (Sekanina 1975). \\
 
This research note was written at the Jet Propulsion Laboratory, California
Institute of Technology, under contract with the National Aeronautics and
Space~Administration.\\[-0.45cm]
\begin{center}
{\footnotesize REFERENCES}
\end{center}
\vspace{-0.55cm}
\begin{description}
{\footnotesize
\item[\hspace{-0.3cm}]
Sekanina, Z.\ 1975, Icarus, 25, 218
\\[-0.74cm]
\item[\hspace{-0.3cm}]
Sekanina, Z.\ 2019, eprint arXiv:1910.08208}
\vspace{-0.1cm}
\end{description}
\end{document}